\documentclass[conference]{IEEEtran}
\IEEEoverridecommandlockouts
\usepackage{cite}
\usepackage{amsmath,amssymb,amsfonts}
\usepackage{algorithmic}
\usepackage{graphicx}
\usepackage{textcomp}
\usepackage{xcolor}
\usepackage{units}
\usepackage[colorlinks=true, urlcolor=blue]{hyperref}

\begin{document}

\title{Enhancing Video Music Recommendation with Transformer-Driven Audio-Visual Embeddings\
}

\author{\IEEEauthorblockN{Shimiao Liu}
\IEEEauthorblockA{\textit{Music Informatics Group} \\
\textit{Georgia Institute of Technology }\\
Atlanta, USA\\
sliu823@gatech.edu}
\and
\IEEEauthorblockN{Alexander Lerch}
\IEEEauthorblockA{\textit{Music Informatics Group} \\
\textit{Georgia Institute of Technology }\\
Atlanta, USA\\
alexander.lerch@gatech.edu}
}

\maketitle

\begin{abstract}
A fitting soundtrack can help a video better convey its content and provide a better immersive experience. This paper introduces a novel approach utilizing self-supervised learning and contrastive learning to automatically recommend audio for video content, thereby eliminating the need for manual labeling. We use a dual-branch cross-modal embedding model that maps both audio and video features into a common low-dimensional space. The fit of various audio-video pairs can then be modeled as inverse distance measure. In addition, a comparative analysis of various temporal encoding methods is presented, emphasizing the effectiveness of transformers in managing the temporal information of audio-video matching tasks. Through multiple experiments, we demonstrate that our model TIVM, which integrates transformer encoders and using InfoNCE loss, significantly improves the performance of audio-video matching and surpasses traditional methods. 

\end{abstract}


\section{Introduction}
Sound has played an integral role for movies for nearly a century.
The 1927 film \textit{The Jazz Singer} marked the beginning of commercial success for films with sound. It is notable for being the first feature-length film with synchronized dialogue and musical sequences, symbolizing the trend of combining music and visual arts \cite{bradley2004first, mundy1999popular}. In films, music is often used to propel the narrative, express character emotions, and enhance immersion. Mundy explains how music intensifies the emotional expression of visual content, deepens the audience's emotional experience, and how visuals and music work together to create a unique viewing experience, similar to the perfect matching of music and choreographed dance movements \cite{mundy1999popular}.

With the rise of video platforms such as TikTok, YouTube, and Instagram, a significant number of users engage with video content daily, either by watching or creating their own videos. Video creators often need to select or produce appropriate background music, as effective background music is crucial for enhancing viewer immersion and aiding the expression of visual content \cite{mundy1999popular}. Selecting suitable background music, however, is not easy, especially for beginners or amateur creators. The music needs to match the video's emotional tone and possibly align with the visual transitions. 
Additionally, users may consider a multitude of other factors, including prioritizing popular or trending songs to increase the potential for exposure. Considering the multitude of factors and choices involved, selecting appropriate music becomes a complex task. This complexity justifies the need for an automated system to assist in making informed choices. It can offer valuable suggestions for users uncertain about music selection, thereby lowering the barriers to video production. Such a system can be implemented on devices that facilitate video production, such as smartphones, computers, VR headsets, and other IOT devices to improve production efficiency by reducing the necessary time for music selection\cite{turchet2023internet}. 


Existing solutions have several limitations. For example, some suggest generating audio based on rhythms or video content. However, audio generated from rhythms may lack musicality due to a limited range of genres and instruments\cite{di2021video}. {Similarly, audio generated based on scene information from video images, such as outdoor environments, is prone to incorporating irrelevant noise and may suffer from poor audio quality \cite{Zhou_2018_CVPR}.} {While many proposed models process features through statistical methods extracted from  videos and audio tracks, they usually do not train using feature sequences that contain rich temporal information \cite{Zhou_2018_CVPR, hong2018cbvmr, zhao2023video}.} Some proposed solutions focus on physically matching audio to video by identifying individuals who are talking or clapping\cite{afouras2020self} or aim at generating foley effects that correspond to real-world sounds such as footsteps or baby cries\cite{9126216,lin2023soundify}, which focuses on a  local instead of global optimization for a necessarily narrow set of scenarios. These approaches do not sufficiently meet  the creative and artistic needs of contemporary video creators. For instance, effective video and audio content often includes segments of buildup and climax, indicating that video and audio sequences are interconnected at various temporal points rather than existing as discrete categories such as train sounds, speech, or baby cries. 

To address these challenges, we aimed to develop a system that recommends music for videos, ensuring that the audio sources are of high quality and rich in variety. Our system recommends entire songs for videos. We conducted the following work:
Firstly, we reimplemented a well-known  model as our baseline and trained it on our data \cite{hong2018cbvmr}. The model leverages self-supervised learning and contrastive learning to recommend audio for video sequences without the need for manual labeling. 
In addition to the Triplet loss and intra-modal structure loss used in the baseline, an additional InfoNCE loss function\cite{oord2018representation}, based on Noise Contrastive Estimation (NCE), is employed. This function effectively learns useful representations from data by contrasting positive and negative samples. It is used to train the model in order to simplify it while maintaining similar or even improved performance \cite{oord2018representation}. 
Secondly, we conducted a comparative analysis of various temporal encoding methods for video and audio sequence features. Our findings suggest that, among the methods we tested, transformers are the most effective for encoding temporal information in audio-video matching tasks \cite{vaswani2017attention}. Our results provide evidence for the critical importance of long-range dependencies in feature sequences for this task. 

\section{Related Work}

\subsection{ Music Video Synthesis}
As early as 2004, Hua et al. introduced an automatic system for generating music videos 
based on onset detection and rhythm estimation of the given audio\cite{hua2004automatic}. However, the generation involved mostly the editing and synchronization of raw video sources with the corresponding music. This method is similar to the method described by Ohya and Morishima in 2013 \cite{ohya2013},  who 
 ``remixed'' existing videos to match the music. Later, Wang et al.\ proposed a system for generating sports music videos, which matched different music segments to video content based on semantic cues such as cheers and goals\cite{wang2005automatic}. This system did not produce a complete musical track but rather a collage of audio clips, and the absence of clear sourcing for the audio materials compromised the audio quality.

\subsection{Music Generation for Videos}
Other methods focus on generating music for existing videos. Zhou et al.\ showed good results for generating waveform samples that are supposed to mimic realistic sounds, such as the rustling of leaves, for videos captured in outdoor environments \cite{Zhou_2018_CVPR}. This method lacked musical components and could generate additional noise.
Di et al.\ introduced a method for generating background music for videos using a controllable music transformer \cite{di2021video}. Trained for six genres and five instruments, this method provides flexibility and control in music generation, however, the audio quality of these MIDI-based creations may not match the richness and fidelity of music produced from high-quality recordings. Furthermore, the constraints on genres and instruments seem to be limiting for generic use cases.

\subsection{Visual-Audio Matching}
For video background music recommendations, Kuo et al. utilized features such as color, texture, lighting, and motion, along with musical characteristics like rhythm and timbre. They applied multi-modal latent semantic analysis to capture correlations between audio and visual terms extracted from these features for music recommendation\cite{kuo2013background}. Hong et al.\ proposed a dual-branch neural network model for cross-modal recommendation between music and videos. They employed both inter-modal ranking constraints such as triplet loss and structural loss to train the model. However, their approach did not incorporate any sequential information over time \cite{cummins2017image}. Van den Oord et al.\ introduced the use of InfoNCE to train contrastive learning models \cite{oord2018representation}. 
Zhao et al.\ proposed a model utilizing multi-level fusion features including video attributes like color, texture, and light, along with music features such as timbre and rhythm attributes to train a convolutional similarity algorithm network for music recommendation\cite{zhao2023video}. Again, it did not utilize sequential temporal information.

Visual-audio matching tasks also involve aligning audio signals with videos. Afouras et al.\ employed self-supervised learning to match audio to videos, which could be used to identify and separate speech from multiple speakers in a video \cite{afouras2020self}. Ghose and Prevost presented ``AutoFoley,'' a method for matching foley sound effects, such as footsteps or clattering objects, to video segments \cite{pretet2021cross}. Similarly, Lin et al.\ presented ``Soundify,'' a system for matching specific sound effects to video scenes, such as the ringing of a bicycle bell or street noise \cite{lin2023soundify}. Their system also automated adjustments to the sound effects' volume and panning to enhance the audio-visual matching. These approaches, however, focus on matching discrete, physical audio cues to corresponding visual events in a segmented manner, such as matching the sound of a baby crying to a visual of a crying baby, or a bicycle bell to the visual of a bicycle. Thus, these methods do not necessarily create a continuous, holistic, and artistically integrated audio-visual narrative. The need for such artistic correspondence in music-video synthesis was highlighted by Surís et al.\ who argue that this correspondence is crucial for enhancing the  overall viewer experience \cite{suris2022s}. However, they did not detail what constituted artistic features beyond stating that these are not merely physical correspondences.

 Our study aims to extend this previous work by conducting a comparative analysis of various temporal encoding methods for video and audio sequence features, with the goal to understand the importance of temporal information in audio-visual matching.

\begin{figure*}
    \centering
    \includegraphics[width=\textwidth]{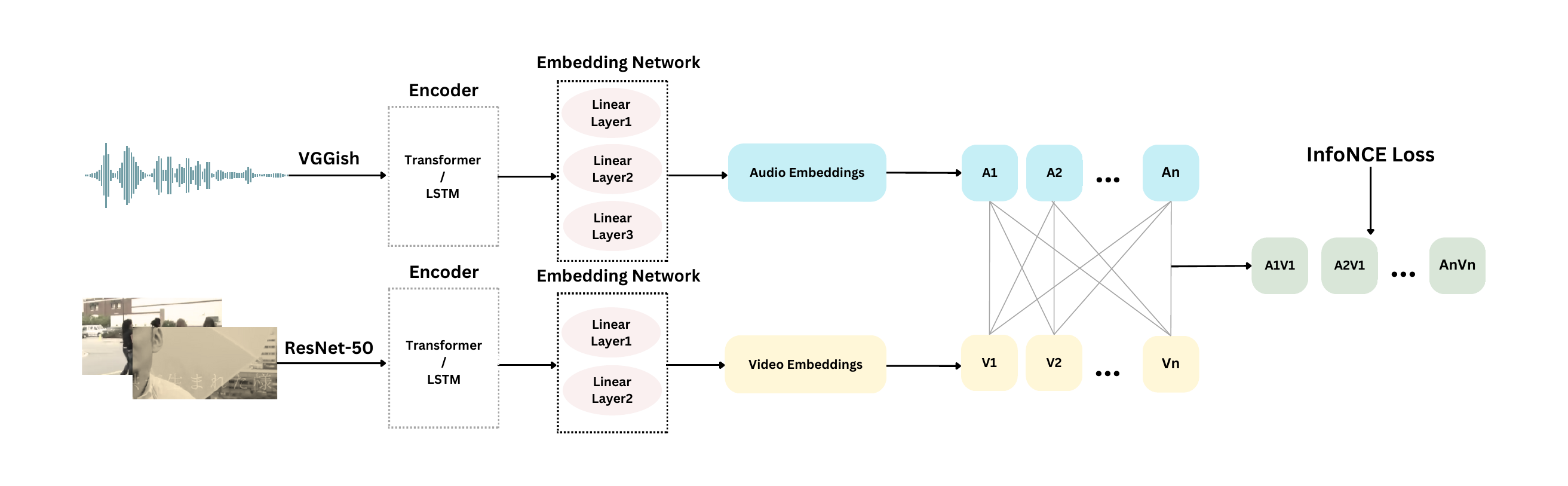}
    \caption{Model Architecture. We use Vggish and ResNet-50 to extract features from audio and video, respectively. The model consists of two separate pathways for processing audio and video, each composed of two or three linear layers. Additionally, we have incorporated an encoder, choosing either a transformer or an LSTM for comparison. The model is trained using the InfoNCE loss.}
    \label{fig:model1}
\end{figure*}

\section{Method}
Our model architecture, as detailed in Figure~\ref{fig:model1}, adapts the VM-NET framework, featuring dual processing pathways for audio and video respectively, each with two to three linear layers\cite{hong2018cbvmr}. We incorporate a choice of either a Transformer or an LSTM as the encoder, and employ the InfoNCE loss for training\cite{oord2018representation},\cite{vaswani2017attention},\cite{hochreiter1997long}.

\subsection{Feature Extraction}

We employ the VGGish and ResNet-50 models for processing audio and visual data, respectively, from selected music videos in the YouTube-8M Dataset \cite{hershey2017cnn},\cite{he2015deep},\cite{abu2016youtube}. We also investigated OpenL3 \cite{cramer2019look} for extracting audio and video features, but found no significant improvement in the results.

\subsubsection{Audio Feature Extraction with VGGish}
  The VGGish model\cite{hershey2017cnn}, specifically designed for audio analysis, is a pre-trained neural network model trained on a large-scale audio dataset that extracts 128-dimensional embedding vectors with a time resolution of \unit[960]{ms}. The VGGish features have been shown adept at encoding various audio patterns for various tasks\cite{hershey2017cnn}. Given that our sample lengths are \unit[15]{s}, these features have the dimensionality (15, 128).

\subsubsection{Visual Feature Extraction with ResNet-50}
For visual data, we utilized the well-known ResNet-50 model\cite{he2015deep}, which is well-known for its effectiveness in image recognition. This model, pre-trained on the ImageNet dataset, extracts detailed visual features from each image to ensure comprehensive coverage of the video content. {For each video, we extracted one frame per second to create an image feature sequence for visual feature extraction.} The ResNet-50 model outputs a 1000-dimensional feature vector for each input frame. The extracted features for our sample lengths result in a video feature dimensionality of (15, 1000) for each input.


\subsection{Baseline Model}
In this study, we reimplemented the dual-branch embedding neural network, VM-NET, to serve as our baseline model \cite{hong2018cbvmr}. Thus, the audio branch has three fully connected layers with node counts of 2048, 1024, and 512. The video branch is implemented with two fully connected layers, with node counts of 2048 and 1024. The network incorporates ReLU activation functions, dropout layers, batch normalization, and L2 normalization. We apply both the cross-modal constraint (Triplet Loss) and the soft intra-modal structural constraint, integrating them with equal weighting to formulate the loss function for the embedding network \cite{weinberger2009distance}. The Triplet Loss aims to minimize the distance between the anchor and positive samples in the embedding space while maximizing the distance to negative samples, with the top 200 most violated cross-modal matches selected for optimization. The soft intra-modal structure constraint aims to preserve the relative distance relationships among samples within the same modality\cite{hong2018cbvmr}. We set $K=10$ for the intra-modal structure constraint, where $K$ controls the number of samples for which the relative distance relationships need to be maintained between the embedding space and the original space.

\subsection{InfoNCE Loss for Cross-Modal Matching}

InfoNCE should be well-suited for contrastive learning as it efficiently handles numerous negative samples within a single batch by treating all other samples as negatives\cite{oord2018representation}. This method typically offers higher computational efficiency compared to configurations that require a predetermined number of triplets in each batch, and it can generate more effective gradient signals. Furthermore, employing InfoNCE can reduce the complexity of the model relative to using a combination of triplet loss and intra-modal structure loss, as it obviates the need for configuring parameters for the top Q most violated matches and managing weighting parameters for two distinct losses. It is defined as:
\begin{equation}
\mathcal{L}_{\text{InfoNCE}} = -\log \left( \frac{\exp(\text{sim}(u, v)/\tau)}{\sum_{k=1}^{K} \exp(\text{sim}(u, v_k)/\tau)} \right)
\end{equation}
where $\text{sim}(u, v)$ denotes cosine similarity, optimizing the model to distinguish between matching and non-matching audio-video pairs effectively. We trained and compared using the InfoNCE loss on top of the baseline model.

\subsection{Encoding of Time Information }

In the two-branch embedding network based on VM-NET we experimented with incorporating an extra encoder for processing temporal sequence information utilizing either an LSTM or Transformer architecture. LSTM, as a specialized type of RNN, can manage sequence information through a series of gates that control the flow of information. These gates include input, forget, and output gates, enabling LSTM to selectively retain or discard information\cite{hochreiter1997long}. Unlike LSTM, Transformers do not process sequences recursively. Instead, they employ a self-attention mechanism that processes the entire sequence simultaneously, allowing for high parallelization and efficient handling of long-range dependencies\cite{vaswani2017attention}.

\subsubsection{{LSTM}}
The model utilizes two bidirectional LSTM layers to process audio and video features independently, with dropout applied to avoid overfitting. The final LSTM output is transformed into a 512-dimensional embedding space using a linear layer, ReLU activation, and dropout.

\subsubsection{{Transformer}}
The Transformer approaches utilizes Positional Encoding to add sequence order information to the input features, crucial for time-series data. The encoder itself then includes multi-head attention and a feedforward network, normalized and regularized with dropout, enhancing the model's capacity to interpret complex dependencies within sequences. The final Transformer output is a 512-dimensional embedding.

\begin{table*}
\caption{Evaluation results with different configurations (in \%)}
\begin{center}
\begin{tabular}{c|c|c|c|c|c|c|c|c}
\hline
\textbf{Metric} & \textbf{Random Result} & \textbf{VM-M} & \textbf{VM-R} & \textbf{VM-MS} & \textbf{IVM-M} & \textbf{IVM-MS} & \textbf{LIVM} & \textbf{TIVM} \\
\hline
\textbf{Accuracy (Top 1)} & 0.06 & 0.37 & 0.20 & 0.49 & 0.55 & \textbf{0.76} & 0.56 & \textbf{2.61} \\
\textbf{Top 5 Recall} & 0.31 & 1.85 & 1.17 & 1.29 & 2.77 & 3.08 & 2.60 & \textbf{12.80} \\
\textbf{Top 10 Recall} & 0.61 & 3.45 & 2.74 & 3.26 & 4.56 & \textbf{5.79} & 4.44 & \textbf{24.01} \\
\textbf{Top 25 Recall} & 1.23 & 7.53 & 6.14 & 7.02 & 9.00 & \textbf{11.58} & 9.26 & \textbf{55.92} \\
\textbf{Top 50 Recall} & 3.07 & 12.52 & 12.53 & 13.61 & 14.73 & \textbf{18.63} & 15.12 & \textbf{98.92} \\
\hline
\end{tabular}
\label{tab:new_results}
\end{center}
\end{table*}

\section{Experimental Setup}
In our experimental design, we investigated various input feature sequence representations including the raw sequence of features, their mean values, and the combination of features means and the standard deviation. The variations also include the different loss function mentioned above and the two temporal encoding architectures. Thus, we arrive at seven distinct experimental configurations.

\subsection{Dataset}
In our research, we have chosen to utilize a subset of the YouTube-8M Dataset for training in audio-video matching tasks. The YouTube-8M Dataset\cite{abu2016youtube} is a vast dataset containing over 6.1 million YouTube video IDs and particularly interests us for its subset of videos labeled as ``music videos,'' as we believe these videos demonstrate a stronger correlation between music and visual content. Compared to purely instrumental videos, such as concert recordings, music videos often provide more complex expressions of content, presumably aligning more with the current trends in video consumption. Thus, we accessed the YouTube-8M Dataset and filtered for videos labeled as ``music videos.'' We then retrieved the full YouTube video IDs for these selected videos and downloaded the corresponding videos along with their audio tracks.

We used 1200 videos labeled as ``music video'' and segmented them into clips of length \unit[15]{s}. This segmentation method aids the training process, as handling longer videos would decelerate both the feature extraction and training phases. Additionally, the \unit[15]{s} length of each segment aligns well with the prevalent duration of short videos today \cite{chen2019study}. After segmenting the videos, we divided them into audio and visual components. This process ultimately produced 16,217 pairs of video-audio suitable for training. These 16,217 pairs of \unit[15]{s} audio-video clips serve as our dataset, which has been divided into training, validation, and test sets in a ratio of 0.8, 0.1, and 0.1, respectively. We ensure that these segments in each set are sourced from different songs. Our dataset is publicly available in our online repository.\footnote{\href{https://github.com/shimiao60s/TIVM}{https://github.com/shimiao60s/TIVM}}


\subsection{Metrics}
We utilize Top $k$ Recall as an evaluation metric. For each sample, a hit is recorded if the positive sample (the original match) is within the top $k$ recommendations. We compute and compare recall at Top 1 (Accuracy), Top 5, Top 10, Top 25, and Top 50 levels. Additionally, we have reported a ``Random Result'' as an anchor, which represents the probability of randomly selecting the correct match from $k$ samples in the test set without using any model. This serves to clarify the effectiveness of the evaluated model.

\subsection{Configurations}
To investigate the performance of our models in video music recommendation and understand the impact of different temporal information processing methods, we designed seven distinct experimental setups. Most configurations represent a different way of handling temporal information.  By comparing the results of these experiments, we aim to identify the optimal method and analyze the underlying reasons.

\subsubsection{{VM-M: VM-NET Using Feature Means}}
We reimplemented VM-NET\cite{hong2018cbvmr} as our baseline model and trained it with our data. In the most basic setup, the input features for audio and video are averaged over time, resulting in input dimensions of 128 and 1000 for each audio and  video snippet, respectively.

\subsubsection{{VM-R: VM-NET with Raw Sequence}}
In this VM-R setup, the entire raw feature sequences of length \unit[15]{s} are the input into the baseline model. We apply max pooling after the embedding process, to facilitate the computation of cosine distances and cosine similarities.

\subsubsection{{VM-MS: VM-NET with Aggregated Features}}
In this configuration, features are aggregated over time to calculate both mean values and standard deviations. Thus, the audio features have dimensions of 256 and the video features have the dimension of 2000. The inclusion of the standard deviation is intended to incorporate more temporal information into the input features.

\subsubsection{{IVM-M: VM-NET Using InfoNCE with Means}}
This configuration retains the dual-branch embedding architecture of VM-NET but utilizes the InfoNCE Loss for training. The input features for both audios and videos are averaged over time.

\subsubsection{{IVM-MS: VM-NET Using InfoNCE with Aggregated Features}}
This model maintains the dual-branch embedding architecture of VM-NET and employs InfoNCE Loss for training. The input features for audios and videos are aggregated over time to both mean values and standard deviations. The addition of standard deviation is to enrich the input features with more temporal information.

\subsubsection{{LIVM: Using LSTM Encoder and InfoNCE}}
This setup preserves the dual-branch embedding structure of VM-NET, uses InfoNCE Loss for training, and incorporates an LSTM Encoder to better manage temporal dynamics. The inputs are raw feature sequences.

\subsubsection{{TIVM: Using Transformer Encoder and InfoNCE}}
The TIVM model maintains the dual-branch embedding structure of VM-NET. It trains using InfoNCE Loss and includes transformer Encoders to enhance handling of temporal dynamics. The inputs are raw feature sequences.

\section{Results and Discussion}

The evaluation results are summarized in Table \ref{tab:new_results}, showcasing the performance of different model configurations across various evaluation metrics.

\subsection{VM-R vs.\ Others}
The baseline using raw sequence as input does not have an advantage compared to VM-M and VM-MS, which use mean values or a combination of mean values and standard deviation as features. Without proper encoding or processing of temporal data, the model's performance is unlikely to improve. Moreover, compared to LIVM and TIVM models, which also use raw sequence as input, the baseline shows no advantage. This indicates that the VM-NET model itself is not suitable for handling sequential information. In contrast, our TIVM model demonstrates strong performance.

\subsection{IVM-M vs.\ VM-M and IVM-MS vs.\ VM-MS }
These two comparisons demonstrate the superiority of InfoNCE. With the same input and VM-NET model, using InfoNCE to train the model yields better results. This improvement indicates that InfoNCE enhances the model's ability to distinguish between matching and non-matching audio-video pairs. When using Triplet Loss, the number of triplets is limited and requires configuration—VM-NET achieves this by selecting the top Q most violated matches\cite{hong2018cbvmr}. In contrast, InfoNCE calculates the relative distances of all pairs, bringing samples closer to their positive counterparts and further from negative ones.

\subsection{VM-M vs.\ VM-MS and IVM-M vs.\ IVM-MS vs.\ LIVM }
The standard deviation calculated from feature sequence reflects~---to some extent---~the changes in features over time. The introducing of the standard deviation into the model input improved the recall rate, highlighting the importance of capturing temporal variations for understanding dynamic audio-visual interactions. However, the LIVM model with an LSTM encoder performed just slightly better overall than the baseline model and the IVM-M model, and did not achieve the same level of improvement as other models with standard deviation input. This suggests that while LSTM can handle temporal dynamics to some extent, it may require a larger dataset or possibly longer sequences to fully realize its capabilities.

\subsection{LIVM vs.\ TIVM}
To increase comparability, we aimed to keep the total parameter count for both LIVM and TIVM within a specific range, such as between 10 million to 20 million. The comparison of Transformer and LSTM enhancements with baseline methods clearly shows that both techniques surpass the baseline across all recall levels. However, significant performance differences exist between them, with the Transformer consistently achieving higher recall rates. This disparity emphasizes the Transformer's superiority in capturing long-range temporal dependencies, essential for tasks involving audio-visual matching. The Transformer’s architecture, capable of processing all elements of the sequence in parallel and its self-attention mechanism, effectively captures complex temporal relationships throughout the sequence. In contrast, while LSTM models are adept at modeling sequential data, their sequential processing nature may limit their ability to handle long-range dependencies. Therefore, the Transformer architecture is better suited for tasks that require a comprehensive understanding and utilization of extensive temporal contexts.



\section{Conclusion and Future Work}
In this study, we have introduced a novel approach to automate the process of recommending appropriate audio for video content, enhancing the effectiveness of audio-video matching. The use of the InfoNCE loss has demonstrated its superiority over traditional loss functions like Triplet loss, enabling our model to better distinguish between matching and non-matching audio-video pairs. Moreover, the implementation of advanced temporal encoding techniques improved the model's accuracy in predictions. Among these, the integration of transformer encoders in our TIVM model has been particularly impactful, effectively handling long-range temporal dependencies essential for the task at hand. Our approach not only surpasses traditional methods but also sets a new benchmark in the field of video music recommendation.

In our future efforts, we plan to introduce more positive pairs, as in actual short-video production, a single song may match multiple videos, or one video may correspond to several songs. This will necessitate adjustments to the dataset or additional manual annotations. We also plan to collect subjective evaluation metrics from users, such as using Pairwise Comparison, to gather user evaluations and rankings of different pairs. This will serve as a criterion to assess the model and incorporate more potentially positive pairs, thereby enhancing our understanding of the model's recommendation results and potentially optimizing the model.


\bibliographystyle{IEEEtran}
\bibliography{Audio-Visual}

\end{document}